\documentclass[10pt]{article}
\usepackage{arxiv}
\usepackage{subfigure}
\usepackage{amssymb}
\usepackage{amsmath}
\usepackage{siunitx}
\usepackage{multicol}
\usepackage{multirow}
\usepackage{algpseudocode,algorithm,algorithmicx}
\usepackage{graphicx}

\newcommand*\Let[2]{\State #1 $\gets$ #2}
\algrenewcommand\algorithmicrequire{\textbf{Precondition:}}
\algrenewcommand\algorithmicensure{\textbf{Postcondition:}}

\usepackage{caption}
\usepackage{url} 

\usepackage[sort&compress,round]{natbib}
\title{Model based Level Shift Detection in Autocorrelated Data Streams using a moving window}

\author{
  Jacob S\o gaard Larsen\thanks{Corresponding author}, Anders Stockmarr, Bjarne Kj\ae r Ersb\o ll, Murat Kulahci\\
  Department of Applied Mathematics and Compute Science\\
  Technical University of Denmark\\
  Richard Petersens Plads, 2800, Lyngby, Denmark\\
  \texttt{\{jasla,anst,bker,muku\}@dtu.dk}
}

\begin{document}
\maketitle

\begin{abstract}
Standard Control Chart techniques to detect level shift in data streams assume independence between observations. As data today is collected with high frequency, this assumption is seldom valid. To overcome this, we propose to adapt the off-line test procedure for detection of outliers based on one-step prediction errors proposed by \citet{tsay1988} into an on-line framework by considering a moving window. Further, we present two algorithms, that in combination, estimate an appropriate test value for our control chart. We test our method on AR(1) processes exposed to level shifts of different sizes and compare it to CUSUM applied to one-step prediction errors. We find that, even though both methods perform comparable when tuned correctly, our method has higher probability of identifying the correct change point of the process. Furthermore, for more complicated processes our method is easier to tune, as the range of window size to be tested is independent of the process.
\end{abstract}

\keywords{Statistical Process Control \and Mean Shift detection \and ARMA process \and Control Chart \and Change Point Detection \and Time Series}

\section{Introduction}
Process surveillance is essential in monitoring the performance of a process over time and discovering special causes that generate more than an expected amount of variation. Process control charts are the main tools in this pursuit of determining whether a process is “in-control”. The classical univariate control charts, also known as Shewhart control charts, are for example based on the assumption of a constant process only exhibiting random variation due to time independent noise, which renders the observations also independent in time. Any deviation from this assumption in the form of a permanent change in the mean (level shift), a spike or a decay (e.g., deterioration due to wear and tear) is expected to be caught by the control chart. Time independence of the observations has long been questioned as data collected in time is expected to exhibit serial dependence, which occurs due to interplay between process dynamics and sampling frequency \citep{bisgaardKulahci2009}. The former reveals the speed with which a process absorbs a disturbance. If sampling is performed faster than the impact of the disturbance on the system is eliminated, the successive observations become correlated as they are affected by the same disturbance that was yet to be eliminated. With the prevalence of automated data gathering systems and advances in sensor technologies, sampling frequency can be made almost arbitrarily high  inevitably yielding serially dependent (autocorrelated) observations. 

There are usually two approaches proposed to alleviate the impact of autocorrelation in the performance of a control chart: (1) model the dynamics out and have a control chart for the leftover, i.e., the residuals, and (2) obtain more accurate estimates of the parameters by taking into account the autocorrelation in the estimation. For the former, the class of autoregressive moving average (ARMA) models is usually employed \citep{boxJenkinsReinsel2008,montgomeryJenningsKulahci2015}. The control chart is then constructed for the residuals as they are expected to be uncorrelated if an appropriate model is used and estimated \citep{alwan1992,alwanroberts1988,montgommeryfriedman1989}.  The first concern in this approach is that the interpretability of the control chart will be negatively affected, as residuals of a model will not convey the same meaning as the original data when plotted on a control chart. But more importantly, it is shown that when residuals of a time series model, such as AR(1) with positive autocorrelation are used for constructing the control chart, the probability of catching a shift in the level is high on the first observation after the shift happens but goes down right after that as the calculations to obtain the residuals adopt to the change in the level \citep{longneckerryan1990,zhang1997}. The second approach also requires fitting a time series model to account for the autocorrelation in the data. Once this is done, then more accurate estimates of the parameters are used to construct the control chart upon which the original data is plotted and monitored. Even if the model fitting were performed successfully, the next concern is that the performance measures of the control chart will be affected when the autocorrelated data is used. The average run length (ARL) is the most commonly used performance measure and refers to the expected number of observations before an out-of-control signal is seen. For an in-control process, ARL is calculated as the inverse of the probability of wrongly labelling a process out-of-control for a given observation. For an out-of-control process, it is defined as the inverse of the probability of correctly declaring an alarm indicating an out-of-control process for a given observation. Both calculations assume independence of the observations. Therefore when data is autocorrelated, the interpretation of the ARL calculated through standard approaches will not be correct. 
 \citet{johnsonbagshaw1974} consider Cumulative Sum (CUSUM) charts, which perform better in detecting small shifts in the mean compared to a Shewhart chart \citep{harris1991}. They propose a formula for adjusting the expected ARL of an in-control AR(1) process. As this method is not suitable for more general ARMA-processes, \citet{aning2017} propose an optimization scheme, that chooses the optimal parameter settings of either CUSUM or Exponentially Weighted Moving Average (EWMA) charts, given an ARMA model and a specified range of interest for the size of a level shift. \citet{zhang1998} proposes the method EWMAST, which uses the autocorrelation function up to M steps to estimate the variation of the EWMA test statistic, and thus correcting the critical value. \citet{apley2002} proposed the Autoregressive $T^2$ Chart, which is a Hotelling $T^2$ chart applied to a moving window of observations from an autocorrelated data series. In order to build the covariance matrix they use the sample autocorrelation function. They conclude that their method works better for a wide range of level shifts and parameter settings of ARMA(1,1) processes than both residual based Shewhart and CUSUM charts. Though, for small level shifts CUSUM is superior, they conclude that the Autoregressive $T^2$ Chart with a fixed window size is suitable for a wide range of level shifts, while CUSUM is very specialized to mean size of interest. \citet{dawod2017} studied the performance of Shewhart X, EWMA and CUSUM Charts when applied to AR(1), MA(1) and ARMA(1,1) processes. They concluded that the best choice of monitoring chart is dependent on the process under consideration. The best choice for AR(1) was CUSUM, for MA(1) it was EWMA and for ARMA(1,1) it was Shewhart X-Chart. \citet{munoz2017} used an Artificial Neural Network to model an AR(1) process and compared the performance of Shewhart X, EWMA and CUSUM Charts when applied to the one step predictions. They concluded that CUSUM was best fitted for this task.

In statistics, outlier detection is used to identify and eliminate the impact of unusual observation in modelling efforts. This is an off-line analysis of data and can be used in Phase I studies in SPC when data analysis is performed for the estimation of the control chart parameters. For real time monitoring (i.e., Phase II), outlier analysis is not suitable without making further adjustments. Based on our literature study we have found that although several specialized methods exists, only the Autoregressive T2 chart by \citet{apley2002} consider the problem as being an online outlier detection problem. We believe this is a very promising approach particularly in cases such as in autocorrelated data. We expect the univariate approach we propose in this paper will have further implications in multivariate settings.

In this article we consider detecting a shift in the level (mean) of serially dependent data. Our approach is based on the outlier detection scheme for time series data proposed by \citet{tsay1988}, which utilizes the structure of the one-step prediction errors to detect level shifts. We adapt this method into a moving window approach, and present two algorithms, that in combination, is efficient at estimating an appropriate critical value for the control chart. The article is organized as follows: In Section 2 we describe our proposed method and algorithms, in Section 3 we describe our test setup. The main results are presented in Section 4 followed by a discussion in Section 5. Finally in Section 6 we present our main conclusions.

\section{Method} \label{sec:method}
We consider the Autoregressive Moving Average (ARMA) model by \citet{boxJenkins} as stated in Eq. \eqref{eq:ARMA-pol}. Here $\Phi(B) =1 - \sum_{i=1}^p \phi_i B^i$ and $\Theta(B) = 1 + \sum_{i=1}^q \theta_i B^i$ with $B$ being the backward operator such that $Bx_t = x_{t-1}$. We also consider the Autoregressive representation \eqref{eq:AR-form} and the Moving Average representation \eqref{eq:MA-form} of the process with $\Pi(B) = \Theta^{-1}(B) \Phi(B) = 1-\sum_{i=1}^{p^*}\pi_i B^i$ and $\Psi(B) = \Phi^{-1}(B) \Theta(B) = 1+\sum_{i=1}^{p^*}\psi_i B^i$. We assume that the model parameters are known and that the process is stationary and invertible.
\begin{subequations} \label{eq:BoxJenkins}
\begin{alignat}{1}
\Phi(B)x_t &= \Theta(B) a_t \label{eq:ARMA-pol}\\
\Pi(B)x_t &= a_t \label{eq:AR-form}\\
x_t &= \Psi(B)a_t \label{eq:MA-form}
\end{alignat}
\end{subequations}
For an AR representation up to degree $p^*$, the optimal prediction of $x_t$ at time $t-1$ is \eqref{eq:one_step_pred}, with the corresponding one-step prediction error defined as \eqref{eq:one_step_err}.
\begin{subequations}
\begin{alignat}{1}
\hat{x}_t &= \sum_{i=1}^{p^*} \pi_i x_{t-i} \label{eq:one_step_pred} \\
e_t &= x_t - \hat{x}_t \label{eq:one_step_err}
\end{alignat}
\end{subequations}

We define $\sigma_a$ to be the standard deviation of the innovations. Matrices are denoted using uppercased boldfaced letters, e.g. $\mathbf{A}$, vectors as lowercased boldfaced letters, e.g. $\mathbf{a}$. $B$ is the backward operator and polynomials in $B$ are denoted using capital Greek letters, e.g. $\Pi(B)$, with parameters denoted as small indexed Greek letters, e.g. $\pi_1$.

\subsection{Phase I procedure}
\citet{tsay1988} considered a general model where an exogenous disturbance $I_t^{t^*}$ impacts as in \eqref{eq:outlierModel}-\eqref{eq:exo_dist} in univariate time series. Here $I_t^{t^*} = 1$ if $t={t^*}$ and 0 otherwise, is the indicator variable for an exogenous disturbance, $f_t$ occurring at time ${t^*}$, and $\tau$ being the size of the disturbance.
\begin{subequations}
\begin{alignat}{1}
x_t &= f_t + \Psi(B) a_t \label{eq:outlierModel} \\
f_t &= \tau \frac{\omega(B)}{\delta(B)} I_t^{{t^*}}  \label{eq:exo_dist}
\end{alignat}
\end{subequations}
This paper is restricted to consider exogenous disturbances resulting in a shift in the mean value of $x_t$. \citet{chenTiao1986} proposed to model this disturbance by setting $\omega(B)/\delta(B) = 1/(1-B)$, i.e. in combination with $I_t^{t^*}$ this produces the step function $S_t^{t^*} = 1$ if $t\geq {t^*}$ and 0 otherwise. Assuming that a disturbance takes the form of a level shift, we can convert \eqref{eq:outlierModel} into AR-form \eqref{eq:lsModel}, and arrive at a model for the behaviour of the one-step prediction error $e_t$, described by the $H(B)$ polynomial.
\begin{subequations}
\begin{alignat}{1}
\Pi(B)x_t &= \tau \frac{\Pi(B)}{1-B}I_t^{{t^*}} + a_t \label{eq:lsModel}\\
&= \tau H(B) I_t^{{t^*}} + a_t \\
H(B) &= 1 + \eta_1 B + \eta_2 B^2 + \cdots = \sum_{i=0}^\infty \eta_i B^i
\end{alignat}
\end{subequations}
Analysing the one-step prediction error in the case of an unobserved level shift, we find the structure outlined in Eq. \eqref{eq:one_step_analysis_1}-\eqref{eq:one_step_analysis_end}.
\begin{subequations}
\begin{alignat}{1}
e_t &= x_t - \hat{x_t} \label{eq:one_step_analysis_1} \\
&\approx \left(\sum_{i=1}^{p^*} \pi_i x_{t-i} + \tau H(B) I_t^{t^*} + a_t\right) - \sum_{i=1}^{p^*} \pi_i x_{t-i}\\
&= \tau H(B) I_t^{t^*} + a_t \label{eq:one_step_analysis_end}
\end{alignat}
\end{subequations}
Note that the expected value of $e_t$ changes according to $H(B)$ after a level shift has occurred
\begin{equation} \label{eq:inno_mean}
\mathbf{E}(e_t) = \begin{cases} 0,\quad t<t^*\\
\tau\ \eta_i,\quad t=t^*+i,\quad i \geq 0
\end{cases}
\end{equation}
The parameters of $H(B)$ are given by the Cauchy product $\Pi(B) * \left(1/(1-B)\right)$. For instance, this means that for an AR(1) process we arrive at $\eta_i=\eta_1$ for $i>1$, i.e. the first one-step predictions after a level shift will have expected value $\tau$ while the subsequent ones will have a constant expected value at $\eta_1\tau$. Based on the model \eqref{eq:inno_mean}, and assuming that the parameters of $H(B)$ are known, an unbiased estimate of $\tau$ at time $d$ given observations up to time $T$ is given by Eq. \eqref{eq:omegaRho} with $\rho^2_{d,T} = \left(1+\sum_{i=1}^{T-d}\eta_i^2   \right)^{-1}$.
\begin{equation} \label{eq:omegaRho}
\hat{\tau}_{d,T} = \rho^2_{d,T}\left(e_d + \sum_{i=1}^{T-d}\eta_i e_{d+i}\right)
\end{equation}
Assuming that no level shift has occurred, we have that $\hat{\tau}_{d,T} \sim \mathcal{N}(0,\rho_{d,T}^2\sigma_a^2)$. Tsay then uses the corresponding pointwise Wald test statistics given in \eqref{eq:tsTestValue}, and a level shift is considered detected if $\max_{1\leq d\leq T} |\lambda_{d,T}| \geq h$, with $h$ being a suitable critical value. 
\begin{equation}\label{eq:tsTestValue}
\lambda_{d,T} = \frac{\hat{\tau}_{d,T}}{\rho_{d,T}\sigma_a} 
\end{equation}
Tsay does not go into further detail about how to choose the critical value, but states that values of 3.0, 3.5 or 4.0 have provided satisfactory results. As we will show later, the critical value is dependent on both the number of samples and the underlying ARMA model.

\subsection{Phase II procedure}
Applying the above procedure in an on-line monitoring scheme would require all pointwise test statistics to be updated when a new observation is available. Thus, as time goes, the computational burden and memory requirements increase. Furthermore, as the number of pointwise test statistics increases, a fixed critical value is not appropriate, as the in-control false alarm rate will not be constant. For these two reasons, Tsays method is not directly applicable in an on-line setting. In the following we describe how we have adapted the method, such that it is usable when monitoring a process.

To convert the method by Tsay into an on-line application, we only consider a moving window of observations. The width of the window is denoted $K$, i.e. we test for level shifts at time points $T-K+1\leq d \leq T$. This is done as a necessity, as a too wide window would have at least the following two downsides: 1) The computational burden may get too big, and 2) The ability to investigate and respond to an event may have vanished if too long time has passed since it occurred. It is obvious that by setting the window size to 1 the proposed method is equivalent to monitoring the one-step prediction errors using a standardized residuals chart. In the case of autocorrelation close to or equal to 0 this method has high power towards detecting large level shifts but low power towards detecting small level shifts \citep{zhang1997}. When increasing the window size, more information about the process is included into the test, and a higher power against detecting small level shifts is achieved. However, this extra power does not come for free, as one has to increase the critical values, thus increasing the response time for larger level shifts. When choosing the window size, one should therefore consider this trade-off and not choose a larger window size than needed to detect the level shift of interest.

Consider now the pointwise test statistics given in Eq. \eqref{eq:tsTestValue}. Given a new observation $x_{T+1}$ and corresponding one-step prediction error $e_{T+1}$ we can update the test statistics iteratively according to \eqref{eq:testUpdate} (note that $\lambda_{T-K+1,T}$ is discarded, as this falls outside the window considered). 
\begin{equation}\label{eq:testUpdate}
\begin{aligned}
\lambda_{d,T+1} &= \frac{\hat{\tau}_{d,T+1}}{\rho_{d,T+1}\sigma_{a}}  \\
 &= \frac{\rho_{d,T+1}\left(e_d+\sum_{i=1}^{T+1-d} \eta_i e_{d+i}\right)}{\sigma_{a}} \\
 &= \frac{\rho_{d,T+1}} {{\rho_{d,T}}}\lambda_{d,T} + \frac{\rho_{d,T+1}\eta_{T+1-d}}{\sigma_{a}} e_{T+1}
\end{aligned}
\end{equation}
This provides a fast means of computing test values when a new observation is available, as one only needs to update the test values. Furthermore, as $\rho_{d,T}$ as it only depends on $T-d$ it is clear that $\rho_{d,T} = \rho_{d+1,T+1}$ and therefore only needs to be computed once for each $d$. Hence, by letting $\mathbf{e}_T$ in Eq. \eqref{eq:e_multi} be the moving window of one-step prediction errors and $\Lambda_T$ in Eq. \eqref{eq:lambda_multi} be the corresponding pointwise test statistics, one can update the test values as the linear system in Eq. \eqref{eq:testUpdate_multi}.
\begin{subequations}
\begin{alignat}{1}
\mathbf{e}_T &= \left[e_{T-K+1},\dots,e_{T}\right]^T \label{eq:e_multi} \\
\Lambda_T &= \left[\lambda_{T-K+1,T},\dots,\lambda_{T,T}\right]^T \label{eq:lambda_multi} \\
\Lambda_{T+1} &= \mathbf{B} \Lambda_T + \mathbf{c}e_{T+1} \label{eq:testUpdate_multi}
\end{alignat}
\end{subequations}

Using this notation an appropriate test statistic to test for a level shift is then the maximum norm of the pointwise test statistics within the current window as given in Eq. \eqref{eq:tsTestStat}.

\begin{equation}
\Lambda = \|\Lambda_T\|_{\infty} \label{eq:tsTestStat}
\end{equation}

\subsection{Choosing the critical value}
Setting the critical value is essential in order to achieve a desired false alarm rate while still being able to detect a certain level shift in the process. We will here outline a procedure to select the critical value for a given window size. 

Using Eq. \eqref{eq:e_multi} and \eqref{eq:lambda_multi}, then from \eqref{eq:omegaRho} and \eqref{eq:tsTestValue} we can write the test values as the linear system \eqref{eq:lambdaLin} which consequently is distributed as \eqref{eq:lambaDist}.
\begin{subequations}
\begin{alignat}{1}
\Lambda_T &= \mathbf{A} \mathbf{e}_T \label{eq:lambdaLin} \\
\Lambda_T &\sim \mathcal{N}\left(\mathbf{0},\mathbf{A}\mathbf{A}^T\right) \label{eq:lambaDist}
\end{alignat}
\end{subequations}
Therefore, as the test values are correlated, a set of $\Lambda_T$ does not correspond to a sequential hypothesis test, hence one cannot do corrections like Bonferoni or Dunn-Sidak \citep{bonferroni1936,sidak1967}. Instead a critical value should be chosen such that for a predefined in-control error rate $\alpha$, the critical values satisfies the following
\begin{equation}\label{eq:gaussInt}
1-\alpha = P(\|\Lambda_T\|_\infty < h) = \frac{1}{\sqrt{(2\pi)^K|\mathbf{A}\mathbf{A}^T|}} \int\limits_{\left[-h,\ h\right]^K}\exp\left(-\frac{1}{2} x^T \left(\mathbf{A}\mathbf{A}^T\right)^{-1} x \right)\,\mathrm{d} \mathbf{x}
\end{equation}
Solving this using numerical integration will be feasible for small values of K, but as the number of nodes needed develops exponentially with the window size $K$, this procedure is infeasible for large values of $K$.

We propose to utilize the statistical properties of the run length distribution, thus enabling us to determine the critical values using simulation. Given a critical value candidate, $h$, we can estimate the corresponding in-control average run length ($ARL_0=1/\alpha$) by performing $n$ repetitions of Algorithm \ref{alg:simLambda}.

\begin{figure}[b]
\centering
\begin{minipage}{.7\linewidth}
\begin{algorithm}[H]
  \caption{}    \label{alg:simLambda}
  \begin{algorithmic}[1]
    \Require{$h$, $\mathbf{A}$, $H(B)$}
    \Statex
    \State Draw $\Lambda_T$ from $\mathcal{N}(0,\mathbf{A}\mathbf{A}^T)$ s.t. $\|\Lambda_T\|_{\infty}<h$
    \Let{i}{0}
      \While{$\|\Lambda_{T+i}\|_{\infty}<h$}
        \Let{i}{i+1}
		\State $e_{T+i} \sim \mathcal{N}(0,\sigma_a^2)$
		\State Update $\lambda_{d,T+i}, T-K+1+i\leq d\leq T+i$ according to \eqref{eq:testUpdate}
      \EndWhile
      \State \Return{i}
  \end{algorithmic}
\end{algorithm}
\end{minipage}
\end{figure}

Assuming that the false alarms are independent, the number of false alarms of a control chart is binomial distributed and the waiting times between events are geometrically distributed. A Maximum Likelihood estimate of $ARL_0$ is then the sample mean. Letting $\bar{x}$ being the mean of $N$ waiting times, a $1-\beta$ confidence interval of the true $ARL_0$ will then be \eqref{eq:ARL0conf}.
\begin{equation}\label{eq:ARL0conf}
\frac{2N \bar{x}}{\chi^2_{1-\beta/2}(2N)} < ARL_0 < \frac{2N \bar{x}}{\chi^2_{\beta/2}(2N)}
\end{equation}
One procedure to select the critical value could be the following:

\textit{Given an ARMA(p,q) model estimated from in-control data and a window size $K$, pick a range for candidate values for the critical value and do $N$ simulations for each candidate value $h_{cand}$ in this range. Finally pick the critical value yielding an average run length closest to the desired $ARL_0$.}

\noindent This procedure has 2 drawbacks. 1: There is the risk of potentially trying a much to wide range of values. 2: For a large window size, it is computationally expensive to do the simulation compared to a small window size. Instead we propose to start with a small window size (e.g. $K=2$), and then sequentially progress for an increasing window size as described in Algorithm \ref{alg:setThresh}, with $\Phi(h)$ being the cumulative distribution of the standard normal distribution. Note that, since having a window size of 1 corresponds to monitoring the standardized one-step prediction errors, one can analytically determine the critical value and thereby have a lower bound on the test value for $K=2$. One thereby avoids performing an a grid search on (h,K) pairs as the statistical properties of one window size is exploited when a larger window size is investigated. Furthermore, the algorithm also utilizes that, as shown in the paragraph below, any two pointwise test statistics will be non-identical, as the correlation between them is less than 1. This means that the conditional variance $\mathbf{V}\left(\lambda_{T-(K+1),T}|\lambda_{T-K,T},\dots,\lambda_{T,T}\right)$ is larger than zero. Hence, when increasing the window size from $K$ to $K+1$, one has to increase the limits of the integral on the right hand side in \eqref{eq:gaussInt} in order to satisfy the identity.	

\begin{figure}[!t]
\centering
\begin{minipage}{.7\linewidth}
\begin{algorithm}[H]
  \caption{}\label{alg:setThresh}
  \begin{algorithmic}[1]
    \Require{Target $\text{ARL}_0$, Window size $K$, Number of repetitions $N$, Significance level $\beta$}
    
 	\Statex
 	\Let{$h_{test}^{(2)}$}{-$\Phi^{-1}\left(\frac{1}{2 \text{ARL}_0}\right)$}
 	\For{$k \gets 2 \textrm{ to } K$}
 	\Let{ARL}{mean of $N$ repetitions of Algorithm \ref{alg:simLambda}}
	 	\While{$ARL_0 > \frac{2\cdot N\cdot \text{ARL}(h_{test}^{(k)})}{\chi^2_{1-\beta/2}(2N)}$}
	 	\State Increase $h_{test}^{(k)}$
 		\Let{ARL$(h_{test}^{(k)})$}{mean of $N$ repetitions of Algorithm \ref{alg:simLambda}}
 		\EndWhile
 		\If{any $\frac{2\cdot N\cdot \text{ARL}(h_{test}^{(k)})}{\chi^2_{\beta/2}(2N)} < \text{ARL}_0$}
 		\Let{$h_{test}^{(k+1)}$}{largest $h_{test}^{(k)}$ satisfying $\frac{2\cdot N\cdot \text{ARL}(h_{test}^{(k)})}{\chi^2_{\beta/2}(2N)} < \text{ARL}_0$}
 		\Else
 		\Let{$h_{test}^{(k+1)}$}{ $\min h_{test}^{(k)}$}
 		\EndIf
 	\EndFor
	\Let{$h_{opt}$}{candidate value minimizing $|\text{ARL}_0 - \text{ARL}(h_{test}^{(K)})|$}
 	\State \Return{$h_{opt}$}
  \end{algorithmic}
\end{algorithm}
\end{minipage}
\end{figure}

\paragraph*{Correlation structure of test statistics when process is in control}
Given a set of test statistics $\Lambda_T$ and a corresponding window size $K$, then the correlation between the test statistics $\lambda_{d,T}$ and $\lambda_{d+t,T}$ for $T-K+1\leq d, d+t \leq T$ is derived in \eqref{eq:testCorr} with $\mathbb{I}(\cdot,\cdot)$ being the indicator function. From this it is clear that if $t > 0$ then $|\mathbf{C}(\lambda_{d,T},\lambda_{d+t,T})| < 1$ since $\eta_0 = 1$.

\begin{subequations} \label{eq:testCorr}
\begin{alignat}{1}
\mathbf{C}(\lambda_{d,T},\lambda_{d+t,T}) &= \mathbf{C}\left(\frac{\hat{\tau}_{d,T}}{\rho_{d,T}\sigma_a},\frac{\hat{\tau}_{d+t,T}}{\rho_{d+t,T}\sigma_a} \right) \\
& =\frac{1}{\sigma_a^2} \mathbf{C}\left[\rho_{d,T}\left(e_d+\sum_{j=1}^{T-d}\eta_j e_{d+j} \right), \rho_{d+t,T}\left(e_{d+t}+\sum_{j=1}^{T-(d+t)}\eta_j e_{d+t+j} \right)  \right] \\
&= \frac{1}{\sigma_a^2}\rho_{d,T}\rho_{d+t,T}\mathbf{C}\left[e_d+\sum_{j=1}^{T-d}\eta_j e_{d+j},e_{d+t}+\sum_{j=1}^{T-(d+t)}\eta_j e_{d+t+j} \right] \\
&= \frac{1}{\sigma_a^2}\rho_{d,T}\rho_{d+t,T} \sum_{i=0}^{T-d} \sum_{j=0}^{T-(d+t)} \sigma_a^2 \eta_i\eta_j \mathbb{I}\left(i , t+j \right) \\
&= \rho_{d,T}\rho_{d+t,T} \sum_{j=0}^{T-(d+t)} \eta_{t+j}\eta_j \\
&= \frac{\sum_{j=0}^{T-(d+t)} \eta_{t+j}\eta_j}{\sqrt{\left(\sum_{i=0}^{T-d}\eta_i^2\right) \left(\sum_{i=0}^{T-(d+t)}\eta_i^2\right)}} < 1
\end{alignat}
\end{subequations}

\section{Test setup} \label{sec:test} 
To evaluate our method, we test it on simulated data. To assess how the performance is relative to standard methods, we have chosen to compare it to the two-sided Cumulative Sum (CUSUM) chart. Below we describe how we have chosen to initialize our charts, how we simulate our data, how we have chosen the the critical value for each window size and how we set up the CUSUM chart.

\subsection{Initialization of charts}
To ensure that our simulations are as true to a real situation as possible, we are not just initializing our test statistics as 0, but close to the true distribution. For our method, the pointwise test statistics are initialized at time $\bar{t}$ according to Eq. \eqref{eq:lambaDist} such that $\|\Lambda_{\bar{t}} \|_\infty < h$. For CUSUM we have implemented a restart procedure. This procedure sets the CUSUM statistics to 0 at time $t=-50$ and update the statistics with simulated in-control one-step prediction errors. If the CUSUM statistics cross the critical value before time $\bar{t}$, the statistics are reset to 0.

\subsection{Simulation studies} \label{sec:DATA}
We perform two sets of simulation studies, both of AR(1) processes with $\phi_1 \in \left[-0.95,\ 0.95\right]$ and $\sigma_a = 1$ and repeated 20,000 times; The first simulation evaluates the out-of-control average run length $ARL_1$ for each value of $\phi_1$. Both our method and the CUSUM chart uses $\bar{t}=0$. The second simulation utilizes that both charts are able to estimate the time point where the level shift occurred. We evaluate the fraction of times the charts signal at $t^* \pm 10$, where $t^*$ is the time point where the level shift occurs. Here both charts uses $\bar{t} = 0$. For both simulations a level shift is added at time $t^*=1$, hence one-step prediction errors are simulated with mean value given by Eq. \eqref{eq:inno_mean}. The size of the exogenous disturbance $\tau$ is parametrized by $\tau = \delta \sigma_a/\sqrt{1-\phi_1^2}$ 
 with $\delta \in \left[0,\ 2.0\right]$. The one-step prediction error is simulated and the pointwise test statistics for our method are updated according to Eq. \eqref{eq:testUpdate} until a signal is detected and the time point is recorded, while we use the update formulas for the two-sided CUSUM chart given in Equations (9.2) and (9.3) p. 418 in \citep{sqcMontgomery7}.

\subsection{Design of our method}
For our method we consider window sizes, $K$, up to 100. For each pair of $\phi_1$ and $K$ the critical value is selected according to Algorithm \ref{alg:setThresh} with $\beta = 0.05$, $N = 21512$ and a target $ARL_0$ of 370.4 corresponding to the usual $\pm 3$ standard deviations of a Shewhart X-chart or residuals chart. We choose the number of repetitions such that the largest margin of the confidence interval \eqref{eq:ARL0conf} is 5, hence a $95\%$ confidence interval for an estimated $ARL_0=370.4$ is $\left[365.5\ 375.4\right]$.

\subsection{Design of the CUSUM chart} \label{sec:setupCUSUM}
To evaluate the performance our method we use a CUSUM chart as a reference method. When designing a CUSUM chart, one is typically setting the slack variable with the objective to search for a level shift change of a certain size. However, as one-step predictions for AR(1) processes will change mean value according to Eq. \eqref{eq:inno_mean} one has to make a design choice, which balances this behaviour. We have chosen two different settings: Setting 1) tune according to the stable level (i.e. setting the slack variable, $s$, to $\tau \eta_1/2$) and Setting 2) based on simulations, pick the optimal slack variable. We test slack variables in the range $\left[0.05 \eta_1,\ 1.5 \eta_1\right]$. The critical value for CUSUM given a slack variable is determined by solving the related integral equations using the function \textbf{xcusum.crit} from the R-package \textbf{spc} version 0.5.4 \citep{spc054}.

\vspace{3cm}
\section{Results} \label{sec:results}
In this section the main results are presented when testing our method and CUSUM on the data described in Section \ref{sec:DATA}. First we will present the results on each method individually, secondly we will compare our method to CUSUM applied to the one-step predictions with the two settings described in Section \ref{sec:setupCUSUM}. \\

\subsection{Applying our method and CUSUM to AR(1)-processes}
\begin{figure}[!htb]
\centering
\includegraphics[width = .45\textwidth]{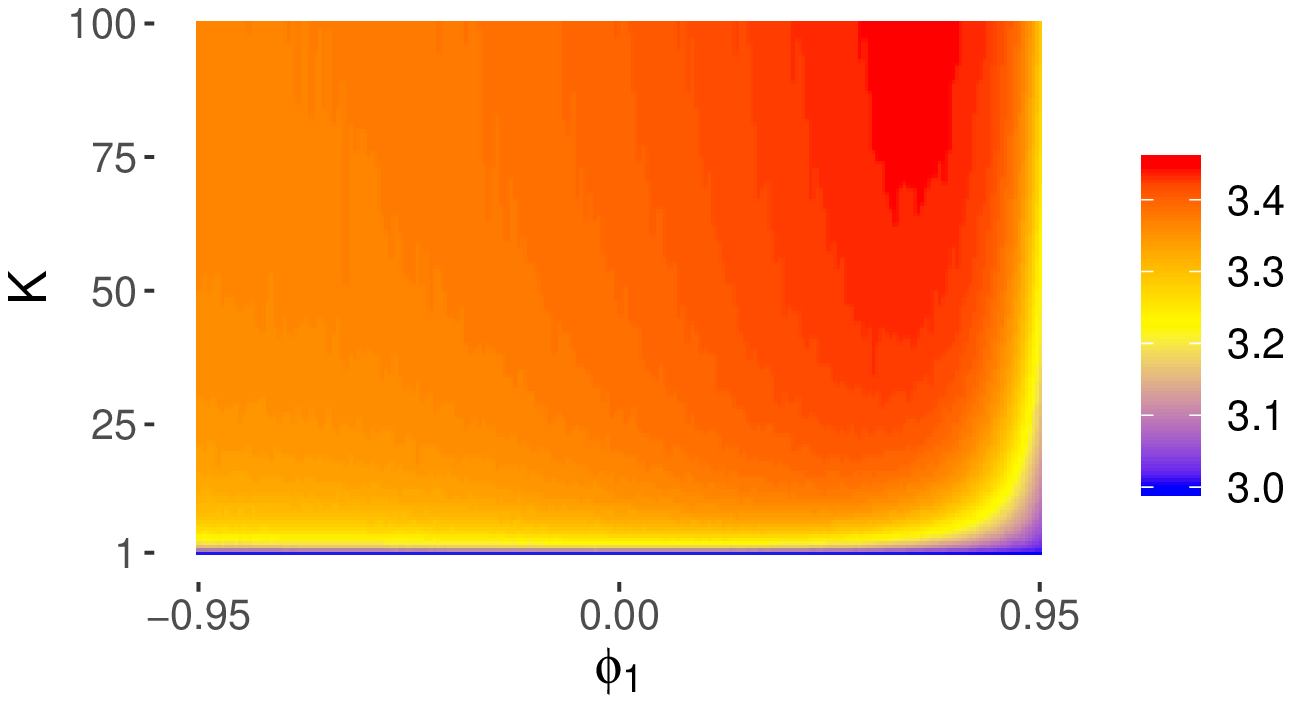}
\caption{Critical values for different window sizes and values of $\phi_1$} \label{fig:filterCrit}
\end{figure}

Figure \ref{fig:filterCrit} shows the estimated critical values using Algorithm \ref{alg:setThresh} for window sizes up to 100 and $\phi_1 \in \left[-0.95\ 0.95\right]$. We see that as $\phi_1$ increases from $-0.95$, the critical value increases as well, but when $\phi_1$ approaches $0.95$, the critical value drops towards 3.

\begin{figure}[p]
\centering
\subfigure[$\delta = 0.5$]{
                \includegraphics[width=.45\textwidth]{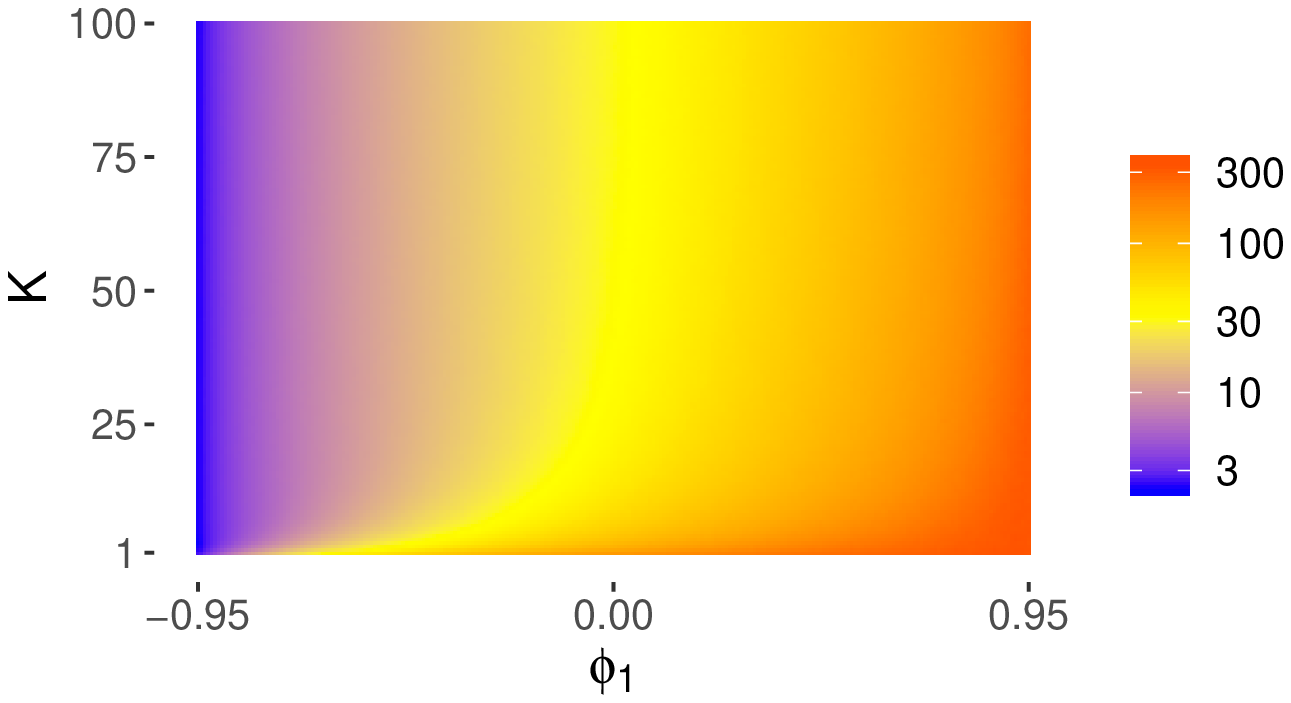}
}
\subfigure[$\delta = 1.0$]{
                \includegraphics[width=.45\textwidth]{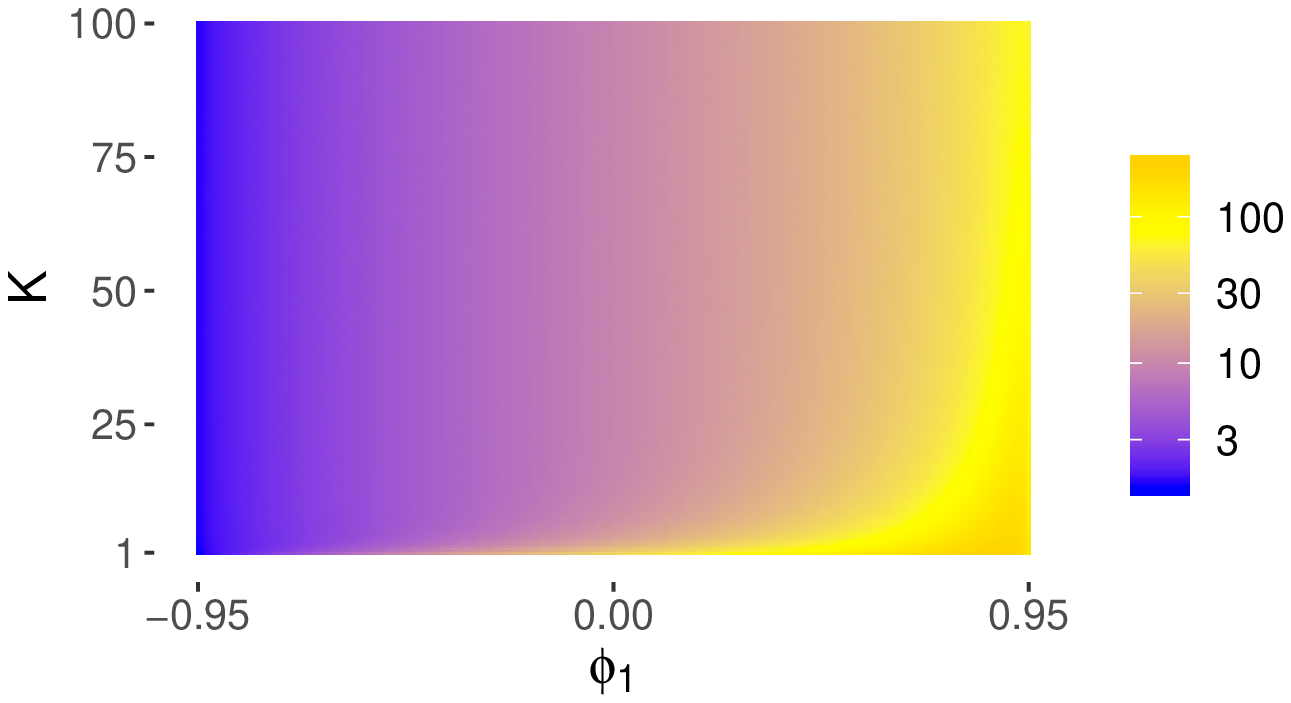}%
}\\
\subfigure[$\delta = 1.5$]{
                \includegraphics[width=.45\textwidth]{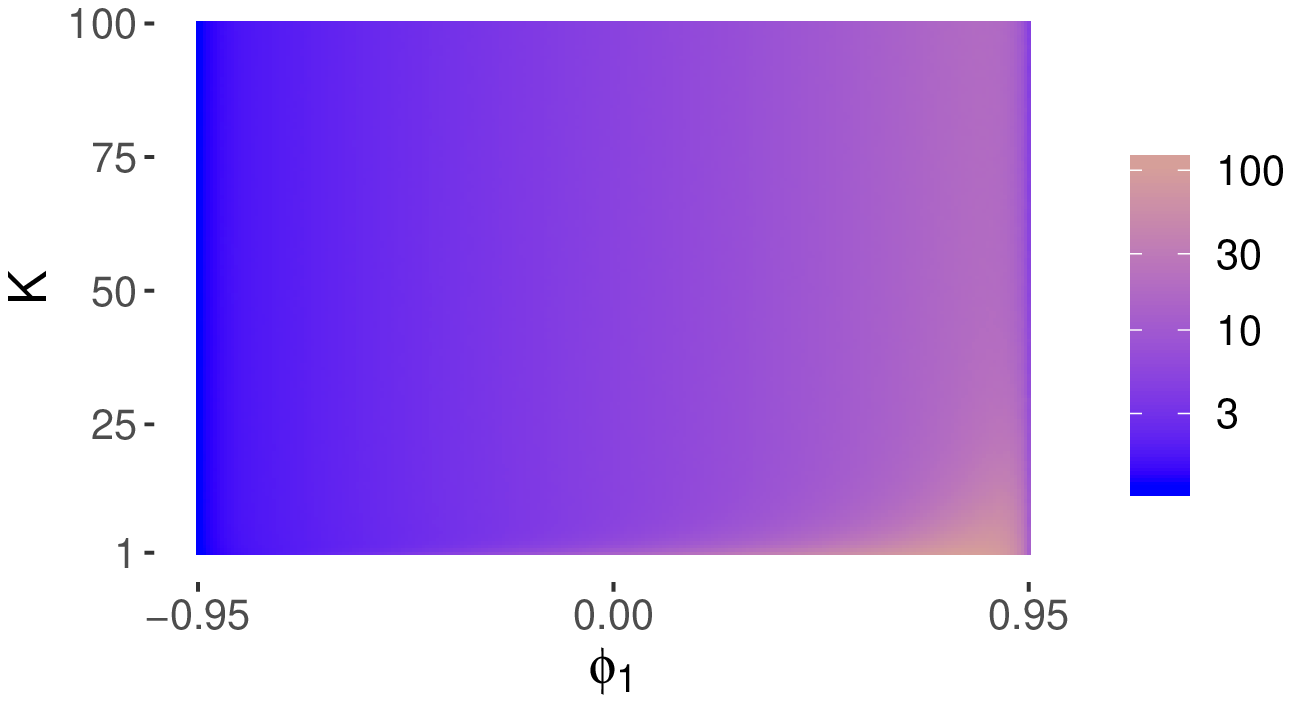}
}
\caption{$\text{ARL}_1$ for our method for three different values of $\delta$.}
\label{fig:filtARL}
\end{figure}

The $\text{ARL}_1$ for our method for 3 different values of $\delta$ and window size up to 100 is shown in Figure \ref{fig:filtARL}. It is seen that as $\delta$ increases, the $\text{ARL}_1$ drops as expected. Furthermore, for $\delta = 0.5$ and $\delta = 1.0$ we see that the $\text{ARL}_1$ increases as $\phi_1$ gets big. For $\delta = 1.5$ $\text{ARL}_1$ first increases when $\phi_1$ is increased from $-0.95$. However, when $\phi_1$ gets big $\text{ARL}_1$ drops. 

\begin{figure}[p]
\centering
\subfigure[$\delta = 0.5$]{
                \includegraphics[width=.45\textwidth]{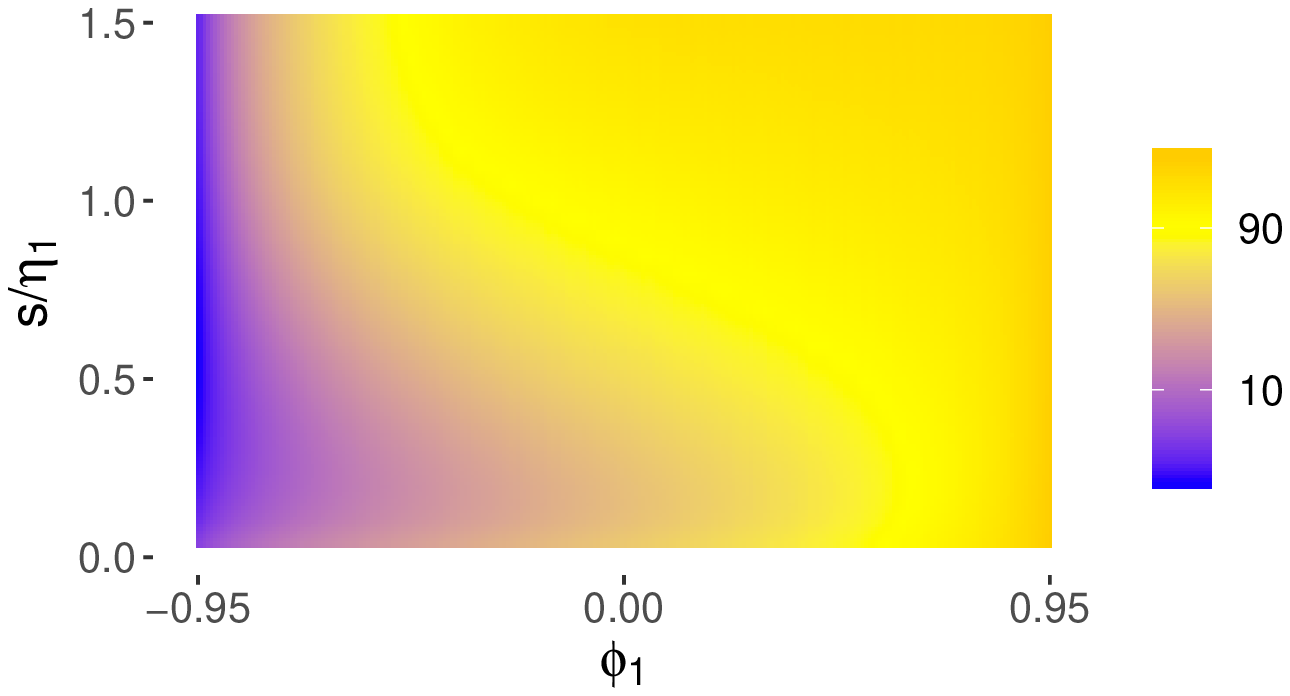}
}
\subfigure[$\delta = 1.0$]{
                \includegraphics[width=.45\textwidth]{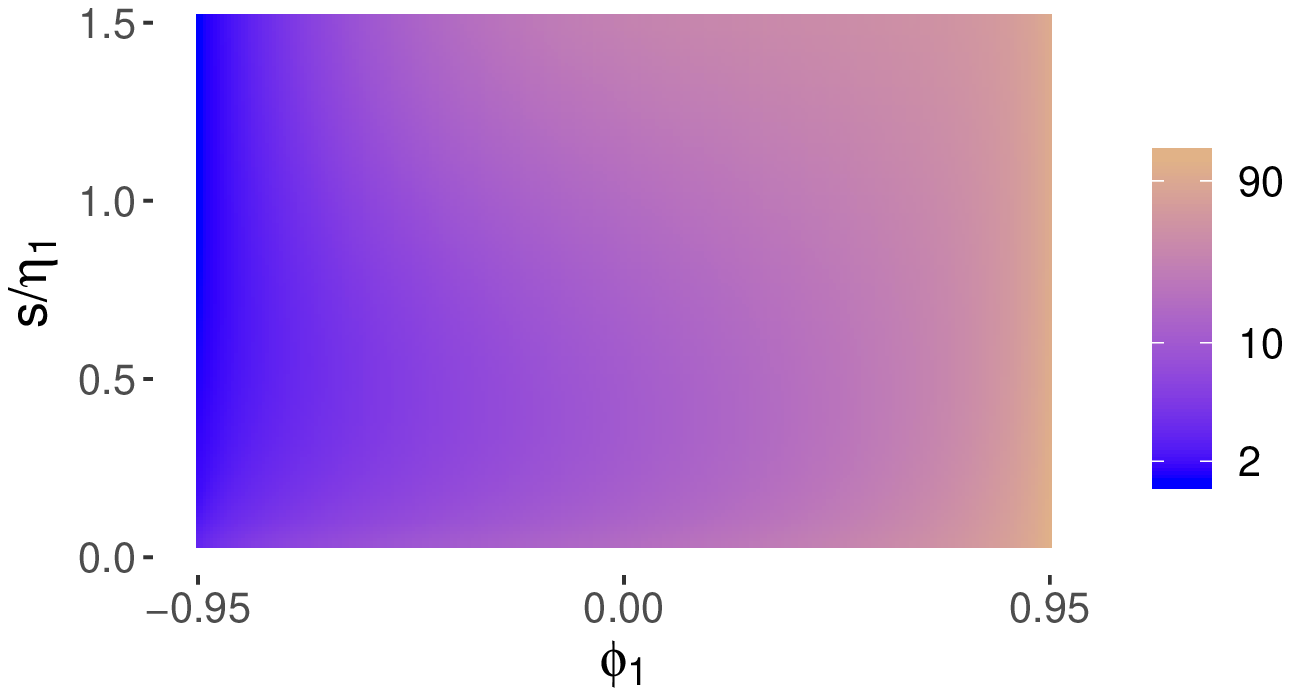}%
}\\
\subfigure[$\delta = 1.5$]{
                \includegraphics[width=.45\textwidth]{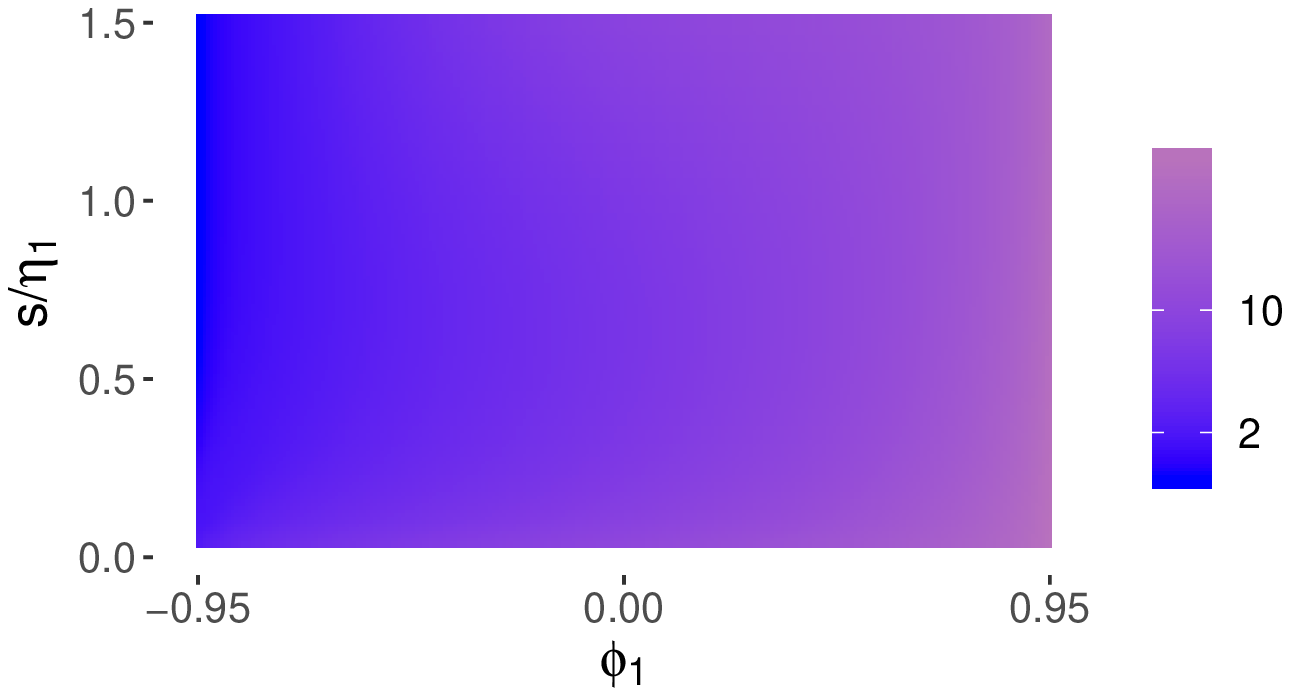}
}
\caption{$\text{ARL}_1$ for CUSUM for three different values of $\delta$.}
\label{fig:cusumARL}
\end{figure}

\begin{figure}[p]
\centering
\subfigure[$\delta = 0.5$]{
                \includegraphics[width=.45\textwidth]{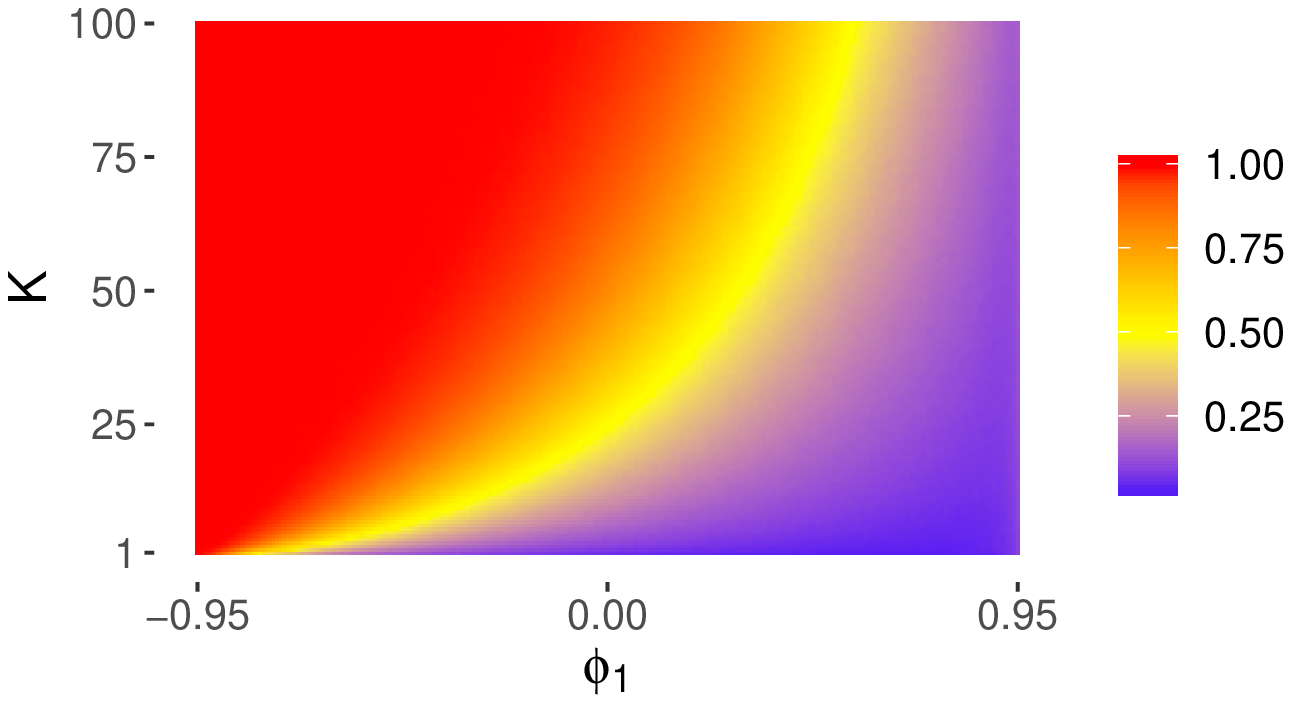}
}
\subfigure[$\delta = 1.0$]{
                \includegraphics[width=.45\textwidth]{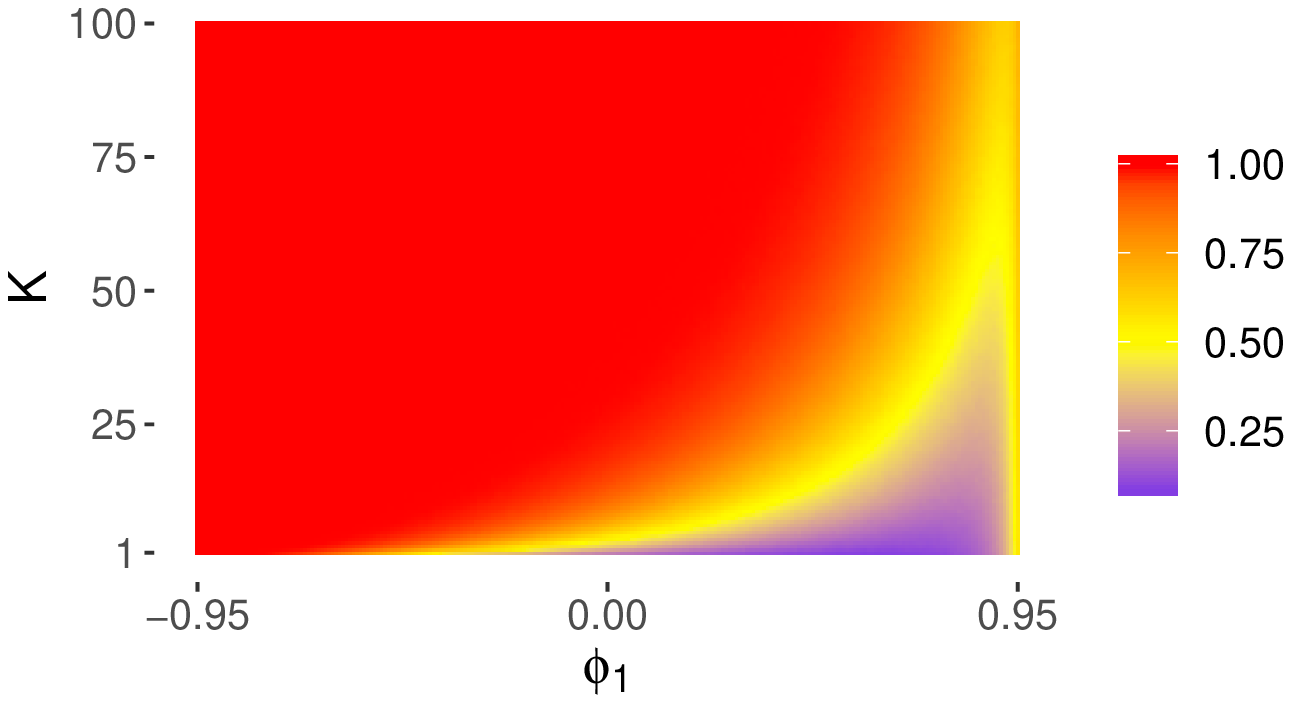}%
}\\
\subfigure[$\delta = 1.5$]{
                \includegraphics[width=.45\textwidth]{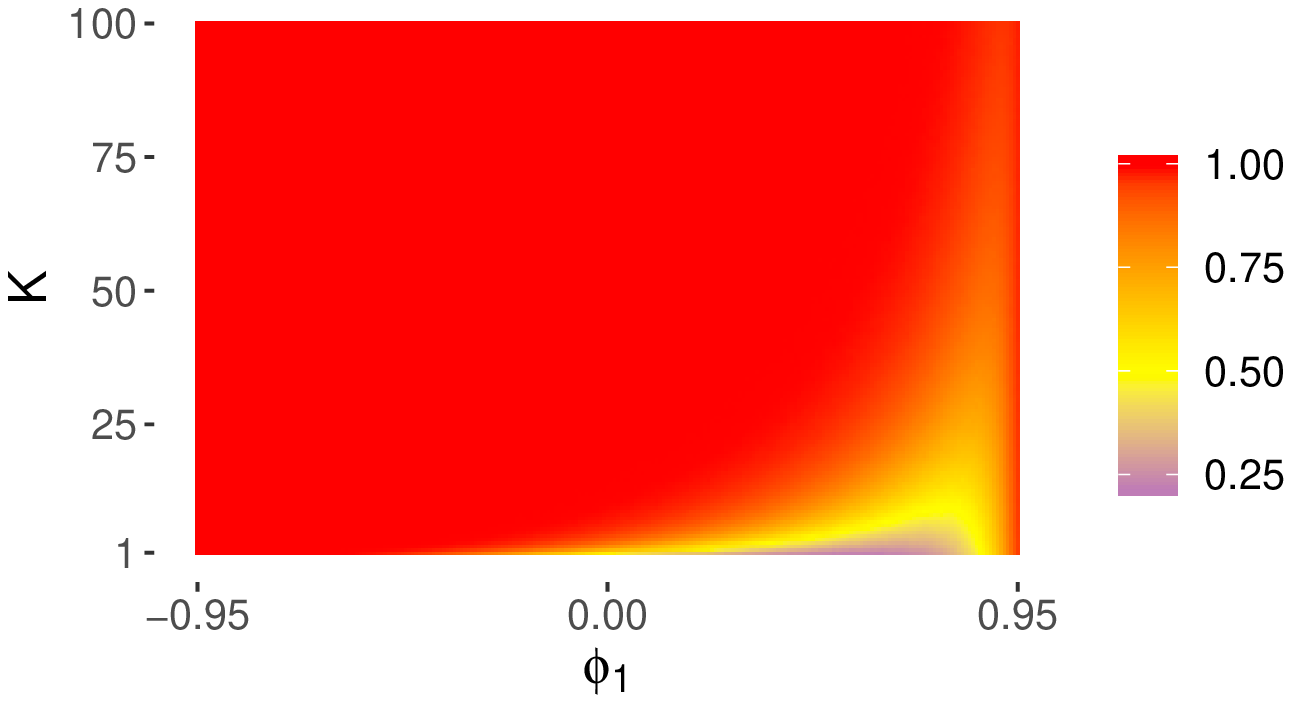}
}
\caption{Probability of signal at $t^*\pm10$ for our method for three different values of $\delta$.}
\label{fig:filtPower}
\end{figure}

\begin{figure}[p]
\centering
\subfigure[$\delta = 0.5$]{
                \includegraphics[width=.45\textwidth]{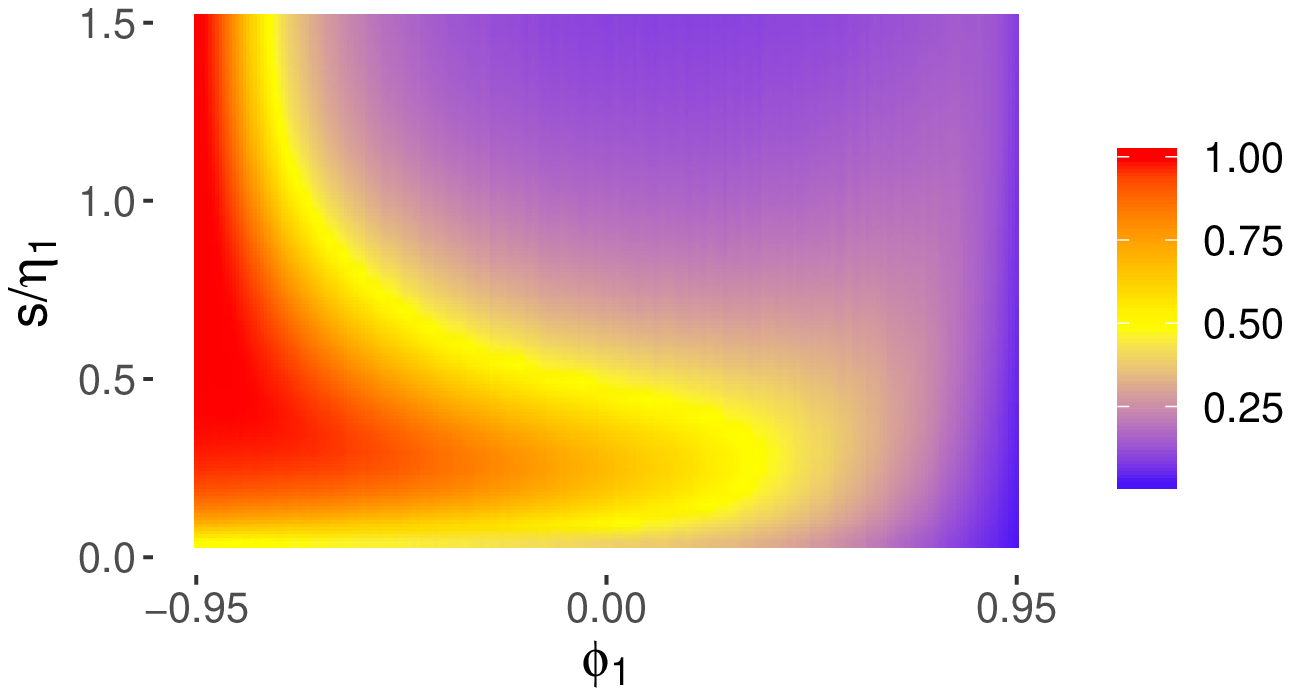}
}
\subfigure[$\delta = 1.0$]{
                \includegraphics[width=.45\textwidth]{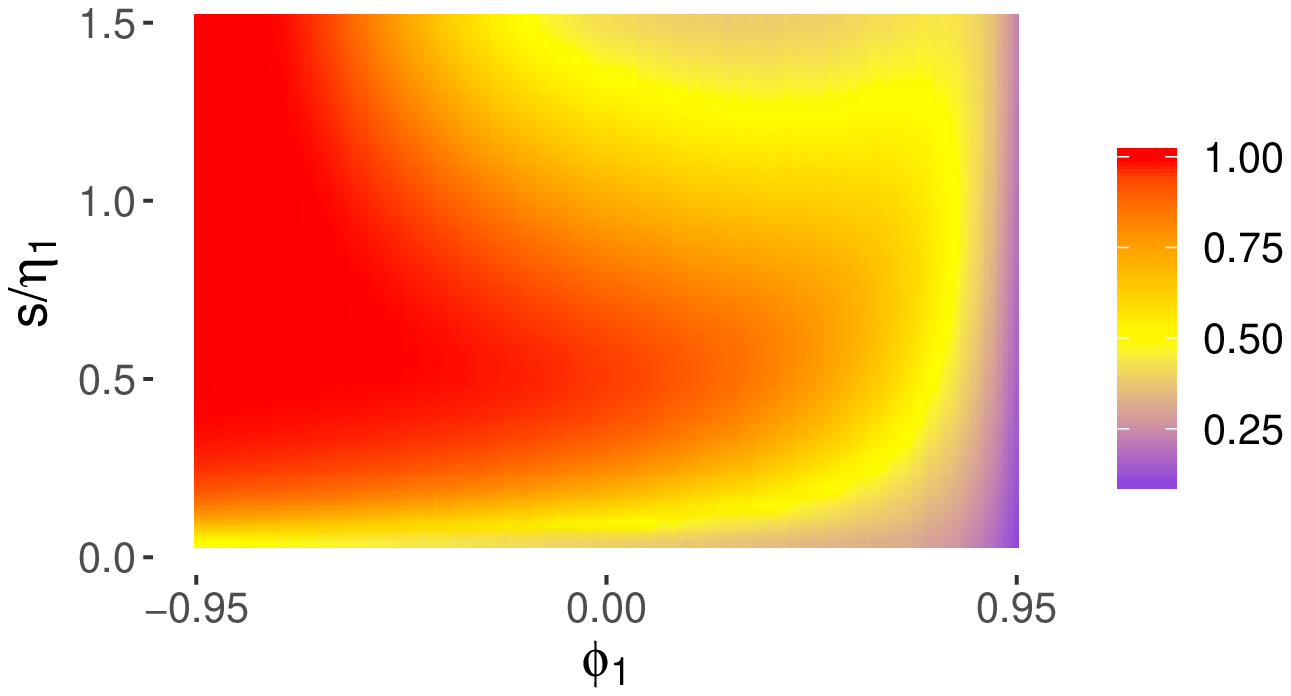}%
}\\
\subfigure[$\delta = 1.5$]{
                \includegraphics[width=.45\textwidth]{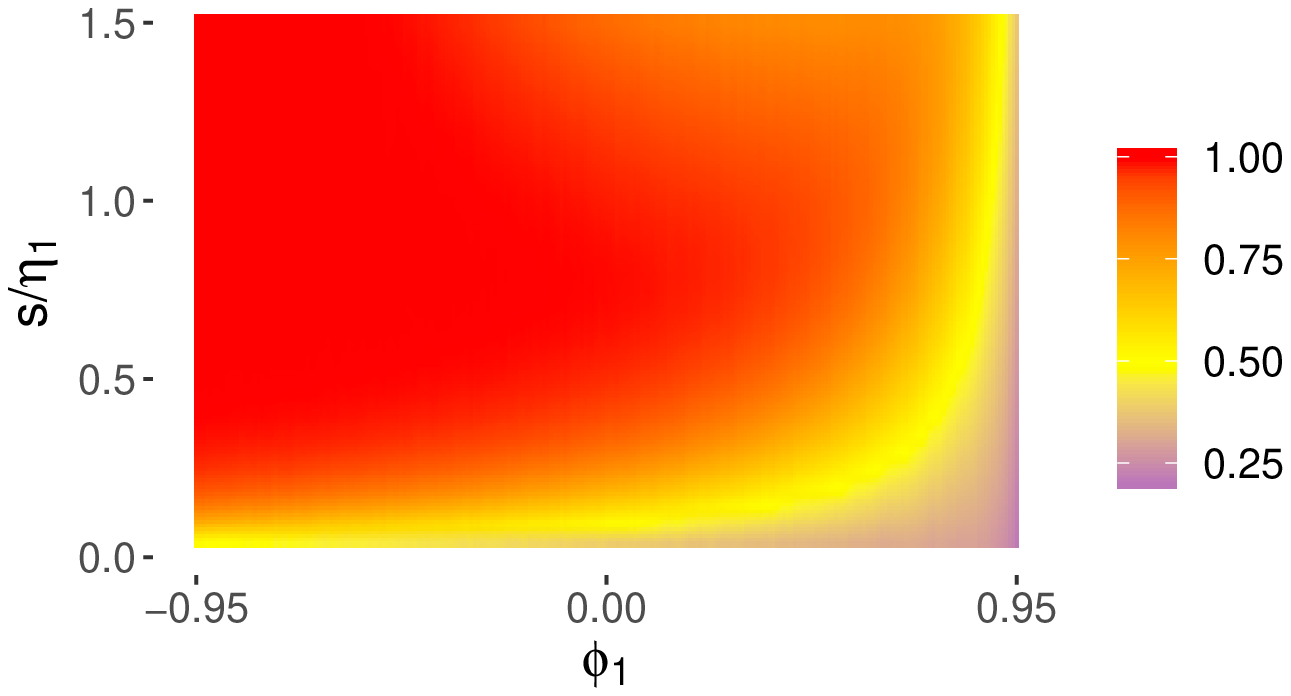}
}
\caption{Probability of signal at $t^*\pm10$ for CUSUM for three different values of $\delta$.}
\label{fig:cusumPower}
\end{figure}

Figure \ref{fig:cusumARL} shows the $\text{ARL}_1$ for CUSUM applied to the one-step prediction errors for 3 different values of $\delta$ and different values of the slack variable. First we note that in both cases, as $\phi_1$ increases from $-0.95$, then $\text{ARL}_1$ increases as well. Further we see that for each pair of $\phi_1$ and $\delta$, if the slack variable is either too large or too small, then $\text{ARL}_1$ increases. I.e. in each case there is an optimal value for the slack variable.

Figure \ref{fig:filtPower} shows the fraction of times our chart signalled a level shift at time $t^*\pm 10$ for three different values of $\delta$. It is seen that when the horizon is increased, the fraction of times the chart signalled at the right time point increases. Figure \ref{fig:cusumPower} shows the fraction of times the CUSUM chart signalled a level shift at time $t^*\pm 10$ for three different values of $\delta$. It is seen that for small level shifts, the ability to detect the correct change point is sensitive to the choice of slack variable.

\subsection{Comparing our method and CUSUM}
A comparison of our method with CUSUM is conducted. As the two methods have different parametrizations, we pick the setting of our method yielding the lowest $\text{ARL}_1$ for each pair of $\delta$ and $\phi_1$. These are compared with the two settings for CUSUM described in Section \ref{sec:setupCUSUM} in terms of the ratio $\text{ARL}_1^\text{CUSUM}/\text{ARL}_1^\text{FILTER}$.

\begin{figure}[p]
\captionsetup{width=0.8\textwidth}
\centering
\subfigure[Setting 1]{
                \includegraphics[width=.45\textwidth]{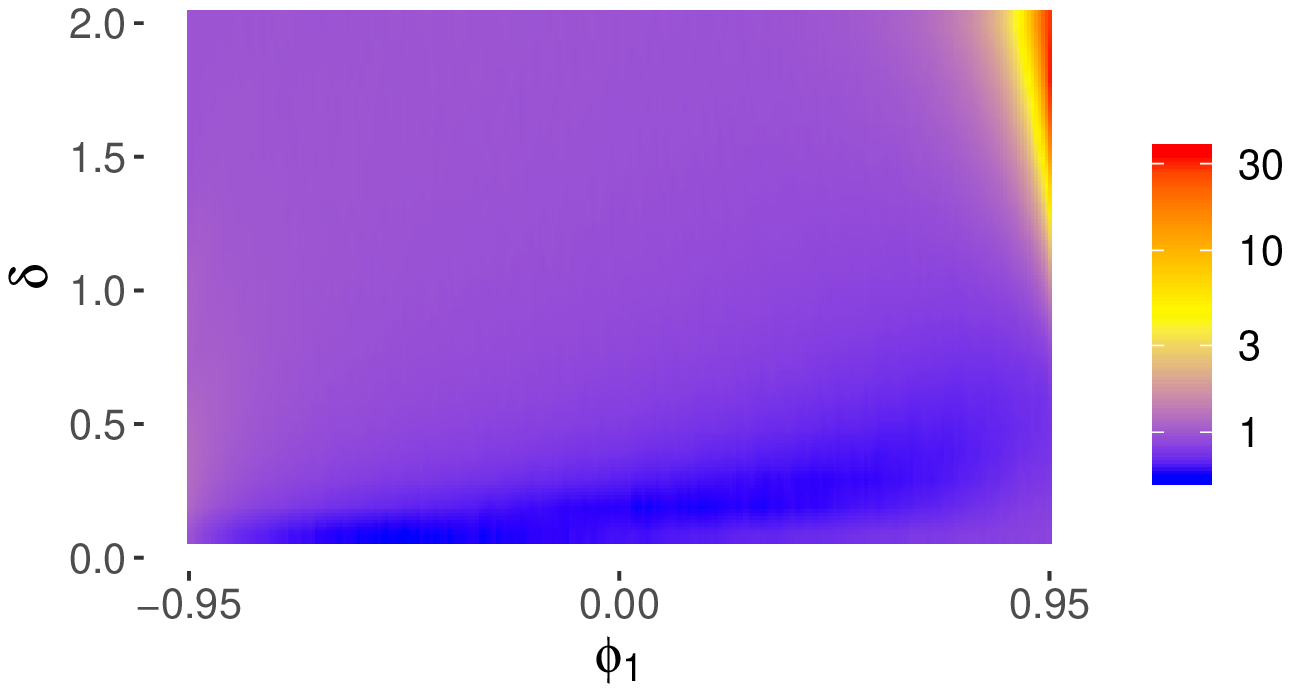}
				\label{fig:STABLE_opt}
}
\subfigure[Setting 2]{
                \includegraphics[width=.45\textwidth]{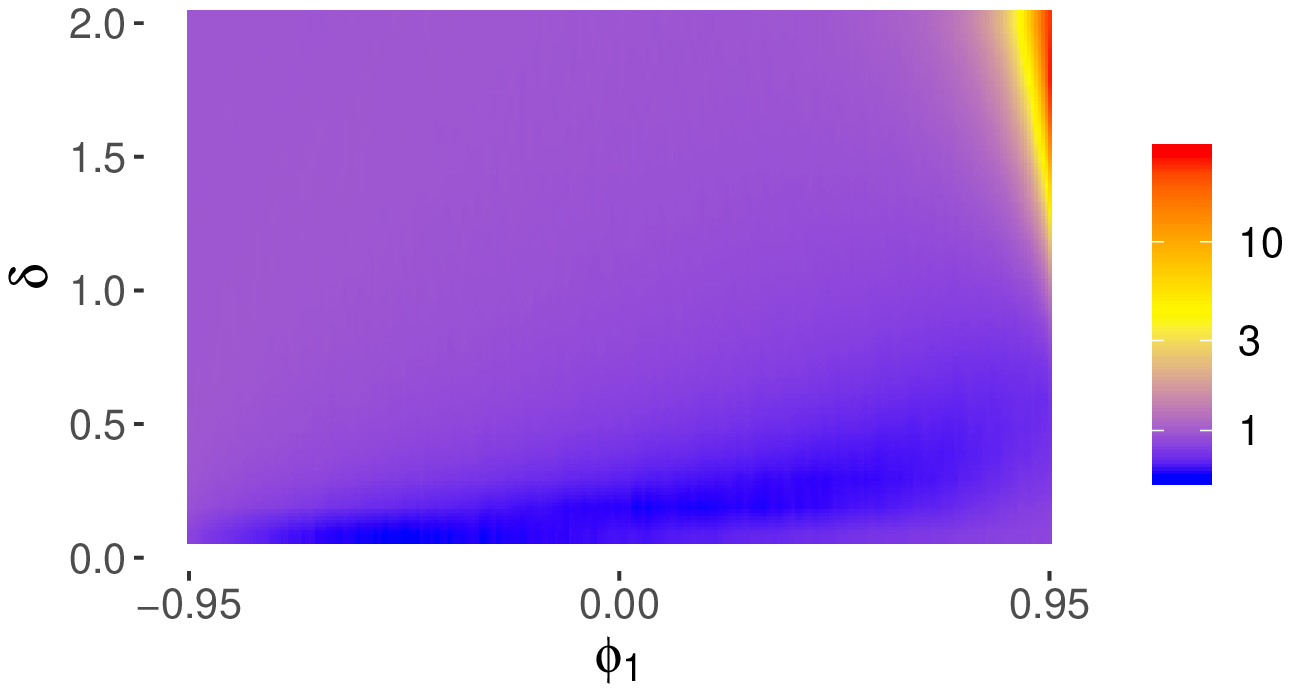}
				\label{fig:OPT_opt}
}
\caption{Comparison of CUSUM and our method for the optimal parameter setting.}
\label{fig:compare_opt}
\end{figure}

Figure \ref{fig:compare_opt} shows the comparison for both settings. First we note, that there is not much difference between the two settings. The biggest difference is present for $\delta>1$ and large negative values of $\phi_1$. We see that most of the area is covered by ratios 
close to 1. Furthermore, for $\phi_1 \geq 0$ our method is generally better than CUSUM for $\delta > 1$, while CUSUM is better for $\delta<1$. For $\phi_1 < 0$ this threshold decreases as $\phi_1$ becomes more negative.

\begin{figure}[p]
\captionsetup{width=0.8\textwidth}
\centering
\subfigure[Setting 1]{
                \includegraphics[width=.45\textwidth]{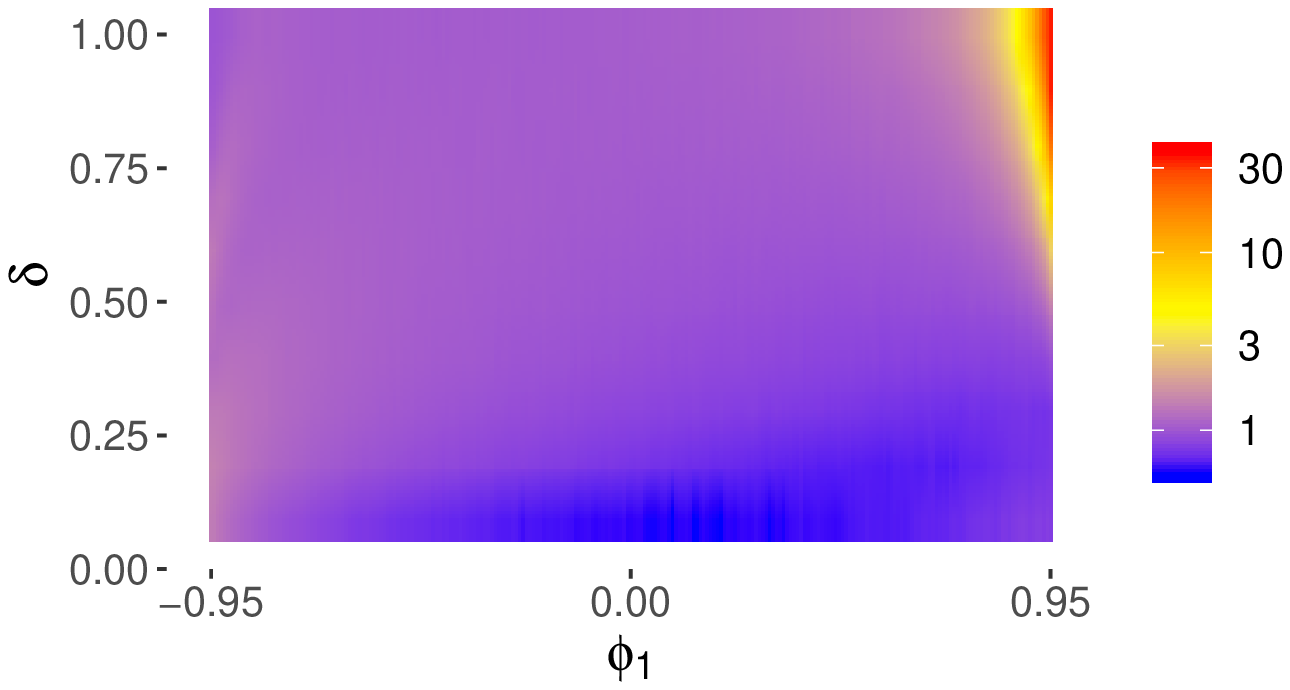}
				\label{fig:STABLE_double}
}
\subfigure[Setting 2]{
                \includegraphics[width=.45\textwidth]{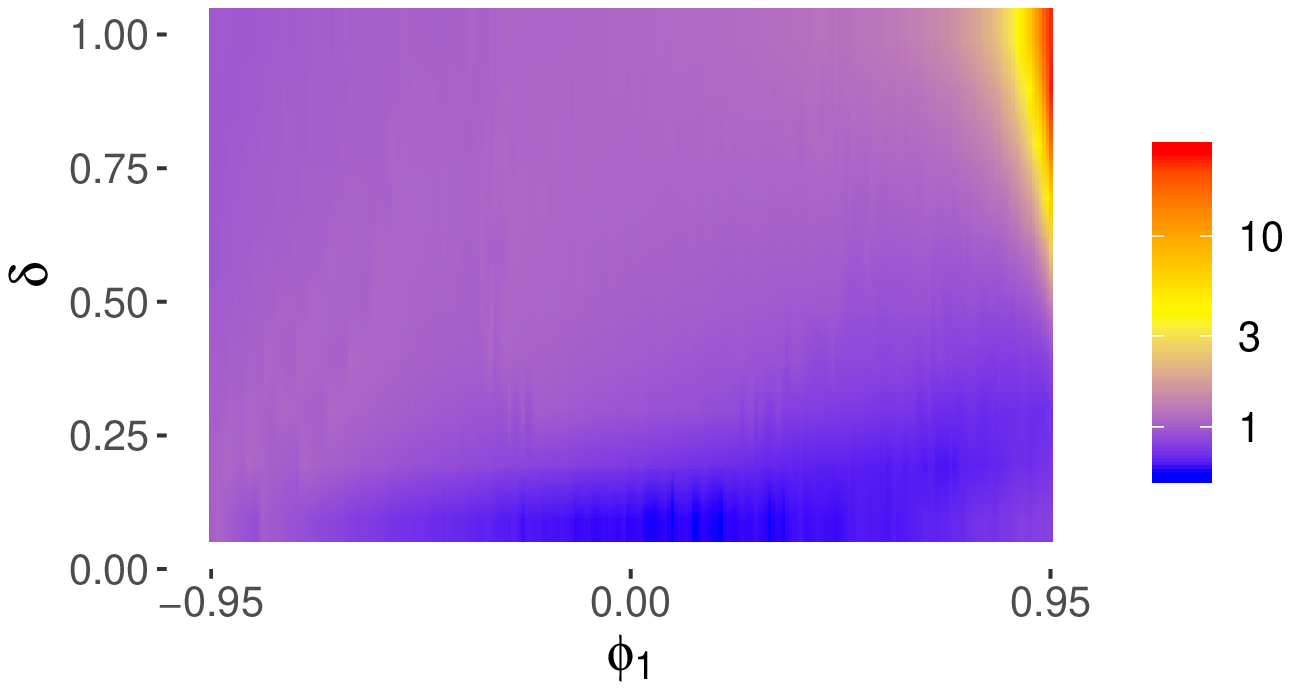}
				\label{fig:OPT_double}
}
\caption{Comparison of CUSUM Setting and our method where the level shift is twice the size as optimal.}
\label{fig:compare_double}
\end{figure}

\begin{figure}[p]
\captionsetup{width=0.8\textwidth}
\centering
\subfigure[Setting 1]{
                \includegraphics[width=.45\textwidth]{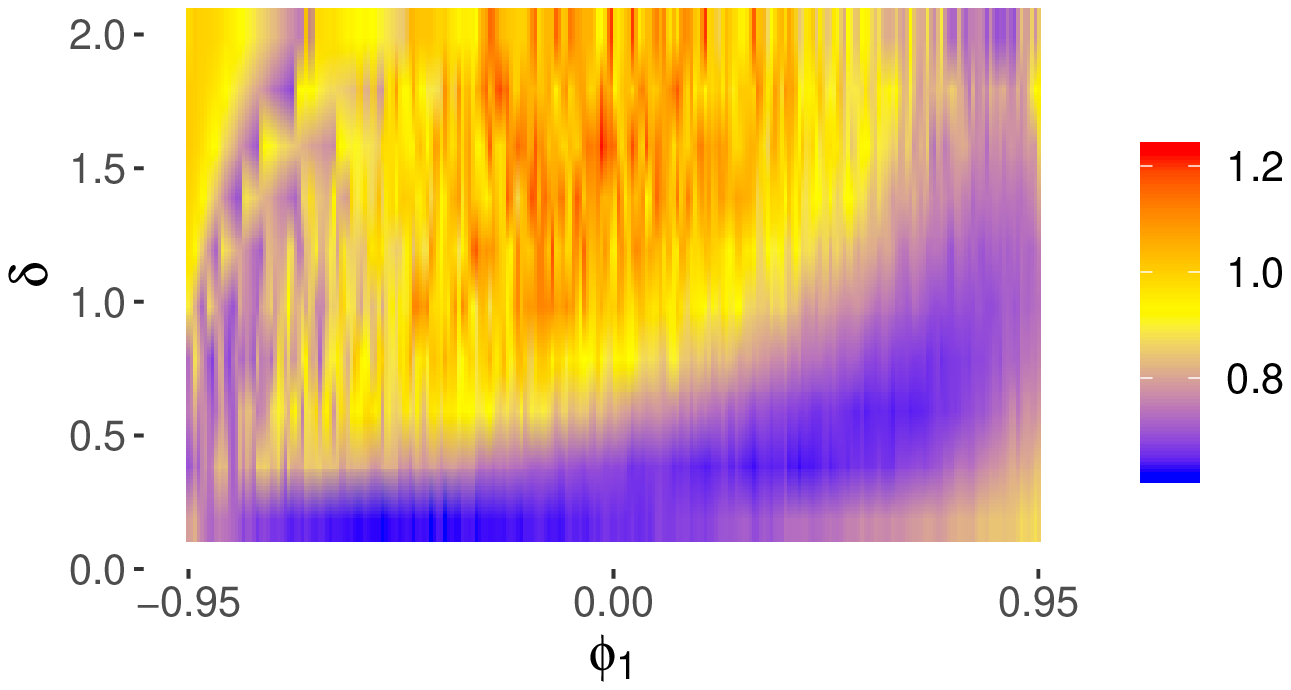}
				\label{fig:STABLE_half}
}
\subfigure[Setting 2]{
                \includegraphics[width=.45\textwidth]{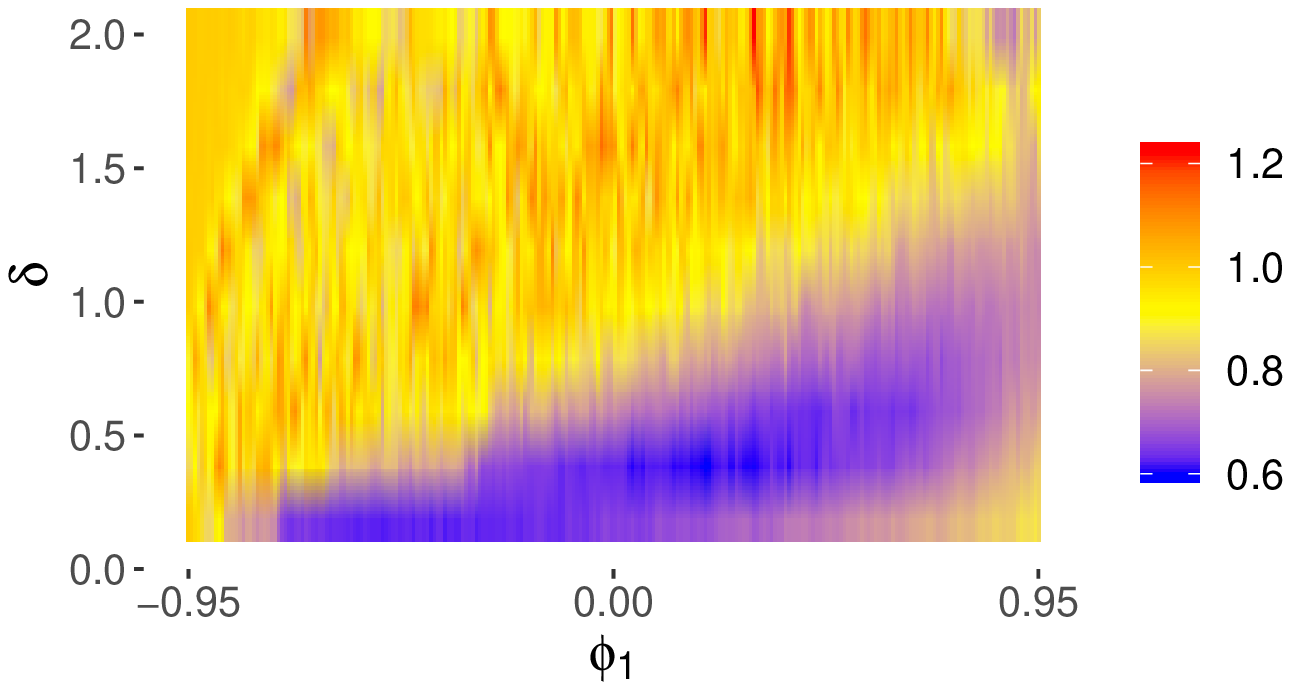}
				\label{fig:OPT_half}
}
\caption{Comparison of CUSUM Setting and our method where the level shift is half the size as optimal.}
\label{fig:compare_half}
\end{figure}

To investigate the robustness towards level shifts of different sizes, we fix the tuning parameters - window size and slack variable - for a level shift of a certain size and analyse how the performance is when either a level shift of half the size or double the size has occurred. The results are shown in Figures \ref{fig:compare_double} and \ref{fig:compare_half}, where $\delta$ indicates the level shift the charts has been tuned for, while the actual level shift is either twice as big or half the size. In Figure \ref{fig:compare_double} the performance of CUSUM and our method are compared when the simulated level shift is twice as large as what the methods are tuned for. As above, there is not much difference between the two settings of CUSUM when compared to our method. It is primarily for large $\delta$ and large negative values of $\phi_1$ there is a difference between the two settings. 
Furthermore, the threshold for which values of $\delta$ our method performs better than CUSUM has been decreased for all values of $\phi_1$

Figure \ref{fig:compare_half} shows the situation where the simulated level shift is half the size of what the methods are tuned for. First we note that it is primarily for either large negative or large positive values of $\phi_1$ there is a difference between the two settings of CUSUM when compared to our method. Here, Setting 1 performs better than our method for all values of $\delta$. For midrange values of $\phi_1$ it is only for small values of $\delta$ that CUSUM performs better than our method. For Setting 2 we can almost draw a straight line from the lower left corner to the upper right corner, where CUSUM will perform better below this line. Above the line, our method generally performs better, however not as uniformly as in Figure \ref{fig:compare_opt}.

Table \ref{tab:comparison} summarizes the results by providing the largest, smallest and median relative differences between our method and CUSUM under both settings and the three different conditions. First it is noted that the ranges do not change between the two settings. However, the median reveals that setting 2 of CUSUM overall performs slightly better than setting 1. Furthermore, there is not much change of the ranges when a level shift of twice the size of the design size is considered.

\begin{table}[b]
\centering
\begin{tabular}{|cc| S[table-format=3.2] S[table-format=3.2] S[table-format=3.2]|}
\hline
\multicolumn{2}{|c|}{Setting} & \multicolumn{1}{c}{Optimal} & \multicolumn{1}{c}{Half} & \multicolumn{1}{c|}{Double} \\ \hline
\multirow{3}{*}{1} & Low & 0.57 & 0.62 & 0.57 \\
& Median & 0.96 & 0.85 & 1.06 \\
& High &  34.44 & 1.23 & 37.46\\ 
\multirow{2}{*}{2} & Low & 0.57 & 0.60 & 0.57 \\
& Median & 0.96 & 0.89 & 1.06 \\
& High & 29.64 & 1.22 & 27.99 \\ \hline
\end{tabular}
\vspace{.1cm}
\caption{Summary of comparison between our method and CUSUM}
\label{tab:comparison}
\end{table}

\section{Discussion} \label{sec:discuss}
%

\subsection{Applying our method and CUSUM to AR(1)-processes}
The critical values increase as expected when the window size is increased. Further, as $\phi_1$ increases from $-1$, then $\eta_1$ will decrease from $2$. This means that the correlation between the test values will decrease. Hence, we will expect that the critical value will increase together with $\phi_1$. Finally as $\phi_1$ approaches 1, a new observed one-step prediction error has decreasing effect when updating the test values as $\eta_1 = 1-\phi_1$.  Hence, at some point this effect will dominate, the method approaches a standardized residuals chart, and a critical value just above 3 is therefore expected when $\phi_1$ approaches $1$.

Our method is seen to have longer $\text{ARL}_1$ when sequential observations are positively correlated. This is expected, as the change in mean value of the one-step predictions ($i>0$ in Eq. \ref{eq:inno_mean}) is small. When approaching non-stationary models $\tau \eta_0$ is large, thus increasing the chance to catch the level shift at the first observation. Furthermore, we note that as the standardized residual chart corresponds to a windoe size of 1, and increasing the window size decreases the $\text{ARL}_1$, our method overall performs better than the standardized residual chart.

CUSUM also shows difficulty detecting level shifts when positive serial correlation is present. Further we observe a sensitivity to correct specification of the slack value. This is expected as CUSUM is known to spend more time catching a large level shift \citep{harris1991}, this corresponding to a slack variable being set too small. Further, a slack variable being set too high decreases the chance that an observation will be larger, therefore increasing the time to signal.

For our method it is seen that the probability of detecting the correct change point is an increasing function of the length of the horizon. The reason for this is that we simultaneously test all time points within our horizon. Naturally, as the correct change point is therefore more likely to be within the horizon when this is increased. CUSUM, on the other hand, is seen to have a range of the slack variable, where it is most likely to detect the correct change point. This range depend both on the underlying process and on the size of the level shift.


\subsection{Comparing our method and CUSUM}
When comparing our method to CUSUM we see that in the chosen range of $\delta$, our method, as seen in Table \ref{tab:comparison} overall performs $39\%$ better than setting 1 of CUSUM, while comparable to setting 2 of CUSUM. We can divide the images in Figure \ref{fig:compare_opt} into two regions, namely for $\delta<1$ and $\delta>1$. For $\delta < 1$ we see that CUSUM has an $ARL_1$ up to $33\%$ lower than our method. For $\delta > 1$ we see that our method performs better than CUSUM. When we consider models with large positive $\phi_1$ our method is up to 48 times better than CUSUM. The reason for this is the behaviour in the expected value of the one-step predictions, as the first will be of the size $\delta \eta_0 = \delta$ and the subsequent of size $\delta \eta_1$, which will be small. This behaviour is directly modelled by our method while setting 1 of CUSUM will be searching for the permanent (small) level shift. For setting 2, the optimal slack value in this region is set to $1.5\eta_1$ (not shown), i.e. the top end of the tested range. This indicates that the optimal strategy in this region will be to search for the first one-step prediction. This will essentially reduce the CUSUM chart to a residuals chart.

When investigating the ability to catch a larger level shift than what the control charts have been tuned to. This effectively means that a too small slack variable has been used for the CUSUM and a too large window size has been used for our method. This means that CUSUM was expected to be affected in terms of a larger $\text{ARL}_1$. In Figure \ref{fig:filtARL} we saw that our method was not affected much if the window size was too large. Hence, in this case we would expect to see a relative improvement for our method compared to CUSUM. 
For setting 2 of CUSUM we overall saw that this was the case, while only the largest relative difference between setting 1 and our method increased. Overall our method still performed better than CUSUM setting 1 as the median was larger than 1.

When the level shift is smaller than what the charts have been tuned to detect, this corresponds to a too large slack variable or a too small window size for CUSUM and our method respectively. In Figure \ref{fig:filtARL} we saw that our method was sensitive to choosing a too small window size. Which is why we see a relative improvement of both settings of CUSUM compared to our method. However, it is often not the concern to catch a smaller level shift than specification, as one could just have decreased the specification prior to tuning.



\subsection{Method performance in the presence of MA-terms}

As we have only investigated AR(1) processes, we have not fully exploited the advances of our method compared to CUSUM. For instance, it is not difficult to find an ARMA(p,q) process where the change in mean value of the one-step prediction errors will behave in a strange way. For instance the first one-step prediction errors after a level shift has occured of the ARMA(2,1) process $x_t -0.4x_{t-1} -0.4x_{t-2} = a_t + 0.8 a_{t-1}$ 
will show a damped oscillation around a constant close to zero. In this case $p^*$ is chosen such that the AR representation fit sufficiently and the horizon is tuned with respect to desired performance in terms of expected $ARL_1$, probability of signalling the correct change point and constraints such as memory. CUSUM, on the other hand, will be more difficult to tune, as some time will go before the expected value of the one-step prediction errors will stop changing sign. We thus expect that studies of our method for more general ARMA processes will unveil significant advantages in terms of parsimony.

\section{Conclusion} \label{sec:conclusion}
In this paper we have proposed to adapt a Phase I procedure for detecting level shifts in ARMA(p,q) time series into an Phase II procedure by considering a moving window, where the size controls the power of the test. Furthermore, based on statistical properties of the method we have outlined two algorithms, that combined, selects the critical value given a specified in-control average run length.

We applied both our method and CUSUM to the one-step predictions of stationary AR(1) processes, exposed to level shifts of different sizes. The performance of the two methods was measured using the out-of-control average run length ($ARL_1$) and the fraction of times the chart signalled a level shift within $\pm 10$ time steps from the correct change point. We saw that CUSUM was more sensitive to a correct parameter selection, as both a too large or too small slack value both increased the $ARL_1$ but also decreased the probability of CUSUM detecting the correct change point of the process, while only a too small window size affected our method significantly.

When comparing the $ARL_1$ of the two methods directly, the two methods performed comparably. Both under optimal conditions, but also when a level shift of a different size than the design specification had to be detected. However, we saw that especially for large $\phi_1$ our method performed much better than CUSUM, while never having an $\text{ARL}_1$ more than $50\%$ larger than that of CUSUM. We saw that it was difficult to design a CUSUM chart to monitor the one-step prediction errors for an AR(1) process. For more complicated process models, this will be even more difficult. Our method does not suffer from this, as the range of the window size to be tested is independent of the process, and thus easier to tune. Further we expect that studies of our method applied to more general ARMA processes will unveil several advantages compared to other methods.

We recommend that future work should investigate the extension to multivariate time series.

\section*{Acknowledgements}
The research is partially funded by BIOPRO (www.biopro.nu) which is financed by the European Regional Development Fund (ERDF), Region Zealand (Denmark) and BIOPRO partners. 

\newpage
{\small
\bibliography{references_1}
}
\end{document}